\documentclass[
  journal=pasa,
  manuscript=research-paper, 
  year=2020,
  volume=37,
]{cup-journal}

\usepackage{microtype,siunitx,booktabs}
\usepackage{gensymb}
\usepackage{hyperref}
\usepackage{xcolor}
\usepackage{soul}
\newcolumntype{P}[1]{>{\centering\arraybackslash}p{#1}}
\title{Image-based searches  for  pulsar  candidates using  MWA VCS data}

\author{S.Sett}
\affiliation{International Centre for Radio Astronomy Research, Curtin University, Bentley, WA 6102, Australia}
\alsoaffiliation{CSIRO Astronomy and Space Science, PO Box 76 Epping NSW 1710 Australia}
\email[S. Sett]{20014515@student.curtin.edu.au}

\author{N.D.R.Bhat}
\affiliation{International Centre for Radio Astronomy Research, Curtin University, Bentley, WA 6102, Australia}

\author{M.Sokolowski}
\affiliation{International Centre for Radio Astronomy Research, Curtin University, Bentley, WA 6102, Australia}

\author{E.Lenc}
\affiliation{CSIRO Astronomy and Space Science, PO Box 76 Epping NSW 1710 Australia}

\received {dd Mmm YYYY}
\revised  {dd Mmm YYYY}
\accepted {dd Mmm YYYY}
\published{22 September 2020}


\begin{document}

\begin{abstract}

Pulsars have been studied extensively over the last few decades and have proven instrumental in exploring a wide variety of physics. Discovering more pulsars emitting at low radio frequencies is crucial to further our understanding of spectral properties and emission mechanisms. The Murchison Widefield Array Voltage Capture System (MWA VCS) has been routinely used to study pulsars at low frequencies and discover new pulsars. The MWA VCS offers the unique opportunity of recording complex voltages from all individual antennas (tiles), which can be off-line beamformed or correlated/imaged at millisecond time resolution. Devising imaged-based methods for finding pulsar candidates, which can be verified in beamformed data, can accelerate the complete process and lead to more pulsar detections. Image-based searches for pulsar candidates can reduce the number of tied-array beams required, increasing compute resource efficiency.  Despite a factor of $\sim$ 4 loss in sensitivity, searching for pulsar candidates in images from the MWA VCS, we can explore a larger parameter space, potentially leading to discoveries of pulsars missed by high-frequency surveys such as steep spectrum pulsars, exotic binary systems, or pulsars obscured in high-time resolution timeseries data by propagation effects. Image-based searches are also essential to probing parts of parameter space inaccessible to traditional beamformed searches with the MWA (e.g. at high dispersion measures). In this paper we describe the innovative approach and capability of dual-processing MWA VCS data, that is forming 1-second visibilities and sky images, finding pulsar candidates in these images, and verifying by forming tied-array beam. We developed and tested image-based methods of finding pulsar candidates, which are based on pulsar properties such as steep spectral index, polarisation and variability. The efficiency of these methodologies has been verified on known pulsars, and the main limitations explained in terms of sensitivity and low-frequency spectral turnover of some pulsars. No candidates were confirmed to be a new pulsar, but this new capability will now be applied to a larger subset of observations to accelerate pulsar discoveries with the MWA and potentially speed up future searches with the SKA-Low.

\end{abstract}

\section{INTRODUCTION }
\label{sec:int}

For decades, searching for pulsars have been performed using time domain search techniques which are highly sensitive for detecting periodic pulses. This approach has been extremely successful and led to majority of pulsar discoveries \citep{ref:ATNF}. However, the sensitivity of these techniques is negatively affected by factors such as binary eclipses, orbital motion and scattering. Especially at low radio frequencies, dispersion measure (DM), smearing and multipath scattering become dominant factors resulting in lesser number of detections. At these frequencies, traditional pulsar searches are also more time consuming and computationally expensive due to the large number of DM trials required at low frequencies. The necessity for performing acceleration searches to detect pulsars in tight binaries also increase the processing time. An alternative strategy to search for bright pulsars is to exploit the high time resolution data of interferometers such as the Murchison Widefield Array \citep[MWA;][]{ref:Tingay, ref:Wayth} to produce continuum images, find pulsar candidates in the image domain and verify them using high-time resolution time-series formed from the very same data. This kind of image-based pulsar candidate searches can potentially reduce the computational resources and processing time when compared to traditional Fourier-domain periodicity searches.\\ 
Many attempts have been made to search for pulsars in radio continuum surveys. Although many of them were unsuccessful, the first ever millisecond pulsar discovered, J1939+2134, was initially identified in radio continuum images as an unusual compact source with a steep spectrum \citep{ref:Backer}. Furthermore, image-based methods have been gaining momentum in the recent years. For example, the discovery of a highly polarised, steep-spectrum millisecond pulsar in a deep image with the Australian Square Kilometre Array Pathfinder telescope \citep[ASKAP;][]{ref:ASKAP} have demonstrated the rising potential of imaging surveys to detect new pulsar candidates\footnote{Continuum sources are considered candidates until confirmed by standard pulsar searches.}. Another recent discovery of a circularly polarised variable radio source in the Large Magellanic Cloud (LMC) shows the potential to detect new pulsars \citep{ref:Wang}.\\
The primary difficulty in the searching for pulsars in image domain is that the observational data are averaged in time and hinder the ability to extract a pulsar profile from time series. The primary challenge, then, is to distinguish interesting pulsar candidates from other unresolved point radio sources. Such differentiation between pulsars and unresolved radio sources can be done using properties that are characteristic to pulsars. \\
One such metric is the steepness of the average spectrum. It has been successfully employed to discover new pulsars that were otherwise missed via traditional pulsar searching techniques \citep{ref:Frail}. This technique has proved to be highly beneficial in targeted observations \citep{ref:Hyman} as well as large sky surveys \citep{ref:2018}. In addition to spectral index methods, \citet{ref:Dai} demonstrates the identification of probable pulsar candidates whose scintillation time-scales and bandwidths are comparable to the time and frequency resolution of the instrument. Another approach is to search for sources that have a high degree of circular polarisation. Previously, a handful of pulsars have been discovered using a combination of spectral properties and linear polarisation \citep{ref:Navarro}, but there has not been large-scale surveys including circular polarisation until \citet{ref:Lenc}. While some steady circularly polarised emission has been seen from sources like flare stars \citep{ref:Lynch1}, chromospherically active binaries like cataclysmic variables \citep{ref:Mutel}, the majority of polarized sources seen in large scale surveys are pulsars \citep{ref:Lenc}. MWA surveys of linearly polarised sources and pulsar searches were also conducted by \citet{ref:POGS, ref:POGS2}. Several other surveys are currently ongoing, such as a search for pulsars associated with polarised point sources using LOFAR, which detected 2 new pulsars \citep{ref:Sobey} and a circular polarisation survey for radio stars with the ASKAP, resulting in 37 pulsar detections, despite them only looking for flare stars \citep{ref:Prit}.\\

In this paper, we present methodologies for finding pulsar candidates in the MWA offline-correlated observations, test their efficiency by detecting known pulsars, and making an attempt to detect new pulsar candidates. Section 2 describes the instrument and the observation details. Section 3 describes the data analysis and the metrics applied to the acquired data. Section 4 summarises the results of the data analysis and discusses the implications for the MWA and other low frequency telescopes such as the MWA.

\section{OBSERVATIONS AND DATA PROCESSING}
\label{sec:obs}

\subsection{Observations}
\label{subsec:obs}

The MWA is a low-frequency Square Kilometer Array (SKA) precursor telescope, operating in the frequency range of 70-300 MHz, located at the Murchison Radio-astronomy Observatory (MRO) in Western Australia. It is the only radio telescope operating at these frequencies able to access the entire southern sky, making it complementary both to similar low-frequency telescopes in the Northern Hemisphere, such as the Low Frequency Array \citep[LOFAR;][]{ref:LOFAR}, as well as high-frequency telescopes in the Southern Hemisphere, such as the Parkes Radio Telescope (Murriyang). Pulsar physics is one of the key science drivers for the MWA \citep{ref:Beard}. Despite being less sensitive than LOFAR, the combination of a large FoV and an extremely radio-quiet environment makes the MWA ideal for undertaking pulsar searches at frequencies below 300 MHz. \\
In this paper we consider archival data collected with the Phase I and II configuration of the MWA which we use to demonstrate image-based methods for pulsar searches. In 2018, the 128-tile array with a maximum baseline of $\sim$ 3km (Phase 1) \citep{ref:Tingay} was upgraded to provide a maximum baseline of 5.3km with 256 tiles (sets of 4x4 cross-dipole antennas). The Phase II upgrade increased the angular resolution by a factor of $\sim$ 2 and the sensitivity by a factor of $\sim$ 4 as a result of reduction in the classical and sidelobe confusion \citep{ref:Wayth}. Originally envisioned as an imaging telescope, requiring only time-averaged tile cross-correlation products ("visibilities"), it was eventually upgraded to enable the capture of raw complex voltages from each tile with the development of the Voltage Capture System \citep[VCS;][]{ref:Tremblay}. The VCS records high-time and frequency resolution voltage data (100 $\mu s$ / 10 kHz), which provides the opportunity to process the data in different ways and maximise data flexibility. Since pulsar flux densities at low-frequencies can vary significantly from day to day, being able to find pulsar candidates in images formed from the MWA VCS data, and vet them by beamforming the very same data is a very powerful technique unique to the MWA.\\ 
The maximum MWA VCS observation duration used here is $\sim$90 minutes and the central frequency ranges from 154 MHz to 185 MHz. Table \ref{Table:1} shows the observations that are processed and analysed as part of this paper. Along with the observation IDs and corresponding project code, details of the telescope configuration at the time of the observation, duration of the observation, the central frequency of the observation and the main application of the observation in this paper are mentioned in the table. \\

   \begin{table*}
      \centering

    $$
        \begin{tabular}{lllllll}
        \hline
\textbf{Observation Name} &  \textbf{Observation ID} & \textbf{Project Code} & \textbf{MWA Phase} & \textbf{Duration (mins)} & \textbf{Frequency (MHz)} & \textbf{Main use}                           \\
        \hline
        
A & 1276619416  & G0071 & PIIE & 90 & 184.96 & Formulating method 1 and 2         \\
B & 1150234552  & G0024 & PI  & 80 & 184.96 & Formulating method 3               \\
C & 1148063920  & G0024 & PI  & 45 & 184.96 & Independent verification of all methods \\
D & 1274143152  & D0029 & PIIE & 20 & 154.24 & Independent verification of all methods  \\
        \hline
    \end{tabular}
    $$
    \caption[]{Observations processed and analysed as part of this paper. PI represents MWA Phase I Array and PIIE represents MWA Phase II Extended Array. The main application for the observations are given in the table. The first 2 observations were mainly used to establish the criteria and the thresholds required to detect highly significant pulsar candidates. The other two observations were used to independently test the methodologies and determine the efficiency of the different methods. }
    \label{Table:1}
    \end{table*}

\subsection{Data processing}
\label{subsec:processing}

\begin{figure}[hbt!]
    \centering
    \includegraphics[width=\textwidth]{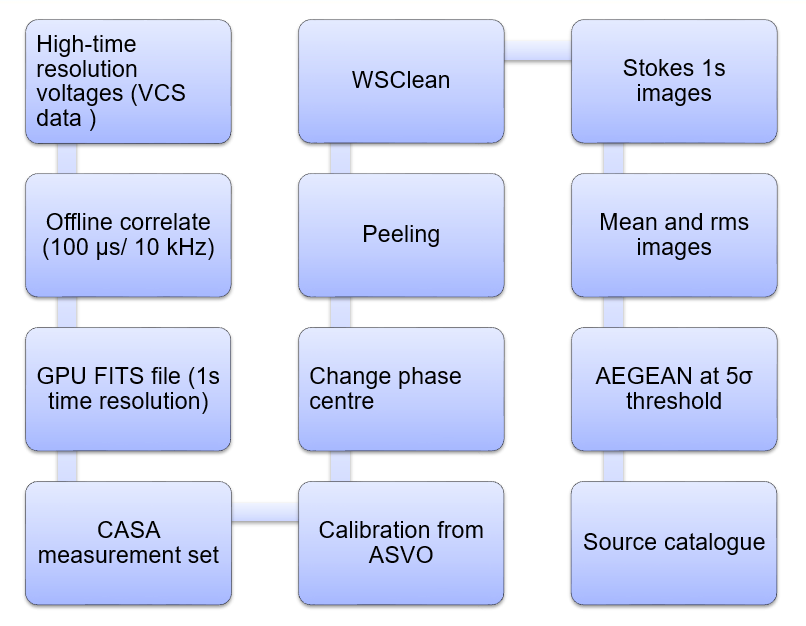}
    \caption{Block diagram of the imaging pipeline, which shows the different steps taken to get the resulting Stokes image. It follows from obtaining the raw voltages, offline correlating and producing measurement sets. WSClean is then used to create the images and apply calibration, ultimately producing Stokes images for every timestamp. These images are then averaged to form the mean Stokes images.}
    \label{fig:bd}
\end{figure}

The raw complex voltages from the MWA antennas (i.e. "tiles") are offline correlated using xGPU software correlator \citep{ref:Clark} to produce "visibilities" at 1-s time resolution. These are then processed with COTTER \citep{ref:Off}, which converts data into the CASA measurement set format \citep{ref:CASA} and applies calibration in the conversion process. It also eliminates the channels affected by radio frequency interference (RFI) by using the in-built software AOFlagger \citep{ref:Offringa}. Calibration solutions for the observation are downloaded from the MWA All-Sky Virtual Observatory \citep[MWA ASVO;][]{ref:Sokolowski}.
Images in instrumental polarisation (XX, YY, XY, YX) are formed using WSCLEAN \citep{ref:wsclean} with a Briggs weighting of -1. These images were then converted to Stokes I, Q, U and V images using the MWA "fully" embedded element beam model \citep{ref:Soko}. The image sizes were 8192$\times$8192 pixels or 4096$\times$4096 pixels, depending on the observation. Individual 1s images were averaged together to produce mean full Stokes images which were then used for further analysis. A block diagram showing the above described steps of the imaging pipeline is given in Figure \ref{fig:bd}. 

\begin{figure*}[hbt!]
    \centering
    \includegraphics[width=\textwidth]{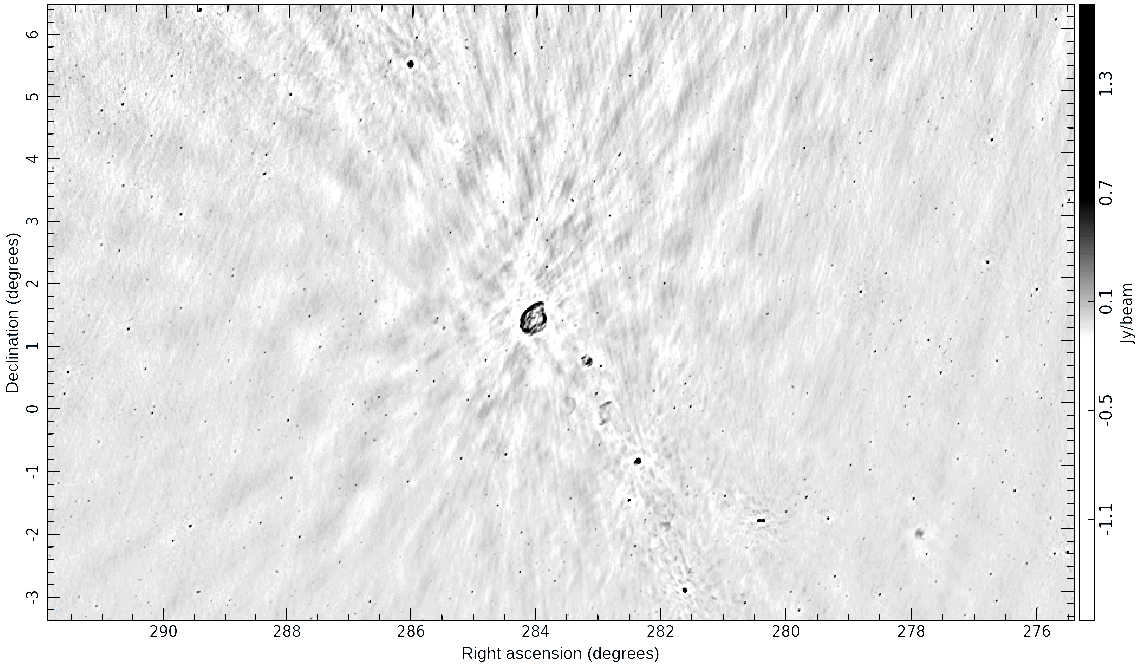}
    \caption{Stokes I mean image cutout of Observation A. It shows the Galactic Plane in the centre of the image. The average rms in the image is $\sim$5 mJy/beam, with a factor of 4 increase near the Galactic Plane and edge of the image. We can also see a variety of extended sources, such as supernova remnants as well as point sources which are a combination of already known pulsars and other radio sources.}
    \label{fig:i}
\end{figure*}

A cutout of a mean Stokes I formed from observation A is shown in Figure \ref{fig:i}. It shows a region of the Galactic Plane passing through the centre of the image along with a supernova remnant which causes significant increase in the noise in that region. The other sources are a combination of stars, active galactic nuclei, known pulsars and other radio sources. The mean rms of this image is 5 mJy/beam and increases by almost a factor of 4 near the Galactic Plane and the edge of the image. Almost 9000 radio sources are detected at 5$\sigma$ threshold in the whole image, where $\sigma$ is the mean standard deviation of the noise. These sources are then used in the methodologies described in Section \ref{sec:methods}. 

The radio source-finding and extraction software, AEGEAN \citep[]{ref:Aegean1,ref:Aegean2} was then used to identify sources from the mean Stokes I and V images and create a catalogue of sources above a 5$\sigma$ threshold which are then analysed to produce a list of promising pulsar candidates. The individual 1s images were also averaged together to produce light curves at various time resolutions such as 300s, 60s and 30s. This can be helpful in detecting sources that are variable in time and hence can be promising pulsar candidates.

\section{Methodologies developed to detect pulsar candidates}
\label{sec:methods}

In the following subsections we describe the methodologies that we have developed and implemented in order to assess the efficiency on known pulsars and also attempt to discover pulsar candidates.

\subsection{Steep spectrum}
\label{subsec:spectral}

Pulsars are known steep-spectrum radio sources, with an average spectral index of $-1.6\pm0.2$ \citep{ref:Jank}. The first millisecond pulsar, PSR J1939+2134 was discovered due to the steep spectral properties of the continuum source, 4C 21.53W \citep{ref:Backer}. The first discovery of a pulsar, PSR J1824-2452A in a globular cluster was made possible due to its steep spectrum \citep{ref:Lyne}. Since then, there have been a handful of target searches aiming to detect steep spectrum sources as pulsars, e.g. \citet{ref:Damico}, \citet{ref:Kaplan} and \citet{ref:Crawford} targeted 18, 16 and 92 sources, respectively, but found no pulsations. While the surveys mentioned above were sensitive enough to detect the associated pulsars with high significance, the available observing frequencies could have affected the detection results for some of the targets. With the advancement in computing resources, such surveys can be re-conducted using techniques such as acceleration searches to detect exotic systems and may be more fruitful. At radio frequencies, the spectral behaviour of many pulsars can be described by a power law, where the spectral index calculated from flux density measurements $\rm S_{a}$ and $\rm S_{b}$, at two different frequencies $\nu_{a}$ and $\nu_{b}$, 

\begin{equation}
    \alpha = \frac{\log \left ( \frac{S_{a}}{ S_{b}} \right )}{\log \left ( \frac{ \nu_{a}}{ \nu_{b}} \right )}.
    \label{equn:spec}
\end{equation}

For our methodology, we have used the Rapid ASKAP Continuum survey \citep[RACS;][]{ref:RACS,ref:racs2} as our counterpart catalogue to calculate the spectral index. The RACS survey is the first large sky survey covering the southern sky below declinations +41$\degree$ using ASKAP \citep{ref:Hotan}. The central frequency of the survey is 887.5 MHz and is one of the deepest radio surveys of the southern sky at these frequencies. The flux densities of the sources from RACS in combination with the flux densities from our images can be used to calculate spectral indices of the sources detected in both images or upper limits on spectral indexes for the sources detected only in MWA image. Since pulsars are known for their steep spectra, it is a compelling case to search for radio pulsations of sources exhibiting a steep spectral index. The results of applying a spectral index cutoff to the sources in the images are described in Section \ref{subsubsec:spec}. Given the angular resolution difference between MWA and ASKAP, cross-matching with RACS is also a good way to eliminate any other extended sources which may have been pre-selected due to lower spatial resolution of the MWA ($\sim$a few arcmin).

\subsection{Circular polarisation}
\label{subsec:polarisation}

Until \citet{ref:Lenc}, most of the all sky surveys have been conducted using total intensity, i.e., Stokes I. As only a fraction of the sources present in the total intensity image emit in circular polarisation, it is useful in lowering the classical confusion limit \citep{ref:Lenc2017}. Greater sensitivity can be obtained for instruments such as the MWA which is confusion limited in total intensity \citep{ref:2013a}. This is extremely helpful for sources that are highly circularly polarised; with a trade-off in sensitivity for sources that are not. The advantage of this is that most sources are generally weak in circular polarisation and hence will not contribute to side-lobe confusion, as a result reducing the need for deconvolution, making image processing easier. A smaller number of sources compared to Stokes I images also reduces the number of candidate pulsars significantly. This method has demonstrated its potential to detect pulsars by the discovery of a highly polarised millisecond pulsar by ASKAP \citep{ref:ASKAP}, which would be otherwise missed by traditional searches due to its wide profile and high DM. We implement the method of looking for sources that are highly polarised in Stokes V images in an attempt to detect any potential pulsar candidates. This method will be highly effective in detecting sources that may be below the noise threshold in Stokes I images but are highly circularly polarised ($>$ 7\%) \citep{ref:Lenc, ref:Lenc2017} and hence very bright in Stokes V images. Further details of this method are discussed in Section \ref{subsubsec:circpol}.

\subsection{Time variability}
\label{subsec:lightcurve}

One of the easiest ways to confirm if a promising candidate is a pulsar or not is to perform a periodicity search and fold the data at the best-fit period and DM. However, at lower frequencies, sources may suffer from significant scattering and dispersion lowers our sensitivity of periodicity searches using high time resolution data. Having images of the field at short timescales for the whole observation opens up a new avenue to detect pulsar candidates. It focuses on how the flux density of the sources vary with time. For example, the ASKAP Variables and Slow Transients (VAST) Pilot project \citep{ref:Banyer} and the LOFAR Transients key science project \citep{ref:Swinbank} have used this methodology to look for transients. Producing light curves essentially requires a priorised fitting for the sources for all the images and construct a light curve from the measured flux density values. This method can be efficient tool for detecting pulsars that may have high degrees of variability on timescale of seconds. The application of this method to the images obtained from our observation is discussed in detail in Section \ref{subsubsec:lightcurve}.

\section{Verification of the methodologies on the MWA data}
\label{sec:results1}

The methodologies described above have been developed and tested on observation A and B (refer to Table \ref{Table:1} for details ). On successful completion of the testing, they have been independently applied to observations C and D (from Table \ref{Table:1}) to verify their efficiency in detecting possible pulsar candidates.\\
\\
We use Stokes I and V images which are of particular interest in the analysis of observation A. The images are 8192 $\times$ 8192 pixels, with pixel size $\sim$ 3 arcmin/pixel, resulting in ~40 $\times$ 40 degree images. The rms noise levels in the image for observation A are around 5 mJy/beam (Stokes I) and 3 mJy/beam (Stokes V) in the centre of both the images. The sources are extracted using AEGEAN \citep[]{ref:Aegean1,ref:Aegean2} with a mean cutoff of 5$\sigma$, where $\sigma$ is the local root-men-square noise level near the source. The mean Stokes I image is complex with over 9000 sources including diffuse emission and several extended structures as shown in Figure \ref{fig:i}. These sources are then used to cross-match with a pulsar catalogue as well as choose promising pulsar candidates, the details of which are discussed in the sections below. \\

\subsection{Noise characterisation}
\label{subsec:noise}

Along with the source catalogues for the Stokes images, AEGEAN also creates maps of the rms noise ($\sigma$) in each image. The $\sigma$ values vary across the image due to a combination of several factors. This includes adjacency to bright sources and unmodelled extended emissions, which are prevalent near the Galactic Plane. The mean $\sigma$ value across the whole field of observation is 5 mJy/beam while it changes to 20 mJy/beam near the Galactic Plane. There is also a significant increase near the edge of the images ($\sim$ 30 mJy/beam) where the beam response of the MWA decreases. Due to this variation of the noise across the whole image, the flux density detection thresholds for the sources should be defined locally and dependent on the location of the source in the image. For example, a 40 mJy source is easily detectable with a 5$\sigma$ significance away from the Galactic Plane. However, the same source will not be bright enough to be detected at the edge of the image where the threshold is significantly higher. 

\subsection{Comparison with the ATNF catalogue}
\label{susbec:atnf}

\begin{figure}[hbt!]
    \centering
    \includegraphics[width=\textwidth]{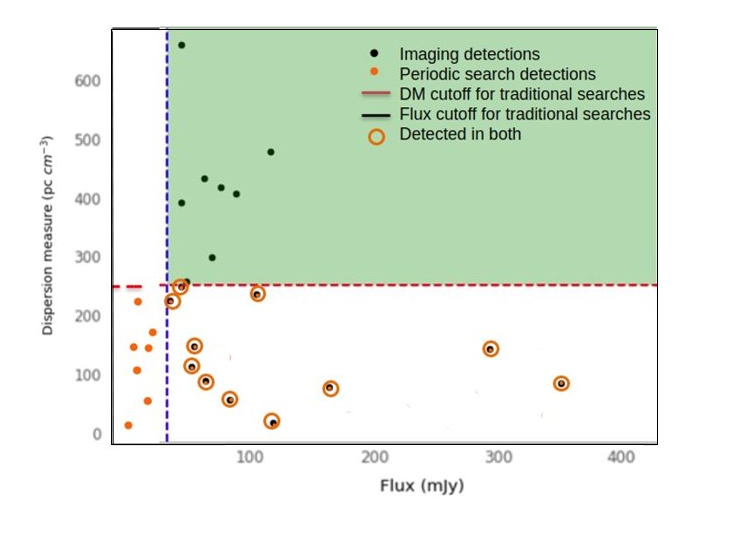}
    \caption{The above figure shows the parameter space that is exclusively accessible to image-based pulsar search techniques, shaded in green. The black dots indicate the image-based pulsar detections and the orange dots represent the detection of pulsars via traditional techniques. The blue dashed line shows the mean flux density threshold of observation A for detection of sources above 5 $\sigma$ (25 mJy). The red dotted line is the DM threshold (250 pc $\rm cm^{-3}$) of traditional search techniques for MWA frequencies beyond which the sensitivity decreases significantly.}
    \label{fig:param}
\end{figure}

The Australian Telescope National Facility (ATNF) pulsar catalogue \citep{ref:ATNF} contains $\sim$ 3000 pulsars. Around 700 pulsars are present in the 3dB (half power point) beam of this observation. We expect to detect 42 of them at 5$\sigma$ threshold when we take into account the local $\sigma$ around the sources. Out of the 42 expected detections, we were able to detect 21 pulsars in imaging which is reasonable considering the more complex environment and the realistic details of the data and processing. Three of the pulsar non-detections can be attributed to their spectral turnover at low frequency as highlighted by \citet{ref:Jank}. The other 18 non-detections were due to the location of these pulsars near the Galactic centre or the edge of the image. The 21 detected pulsars were also folded with PRESTO \citep{ref:Ransom} in order to verify their detections using standard methods. Only 11 out of 21 were detected via PRESTO. This can be explained by the DM cutoff of $\sim$250 $\rm pc ~\rm cm^{-3}$ for the MWA beyond which the detection sensitivity is highly reduced when dealing with time series data. The pulsars detected via imaging (black circles) and traditional search methods (orange circles) are shown in Figure \ref{fig:param}. The blue dotted line indicates the flux density threshold for image based searches for this observation, the red dashed line indicates the approximate DM cut-off for traditional searches with the MWA. The green shaded region highlights where image-based searches can be more successful than traditional searches. In the figure, we can see that even though there are more pulsars detected via traditional searches, image-based searches probe a parameter space that is inaccessible to traditional searches.

\subsection{Comparison with RACS catalogue} 
\label{subsec:RACS}
\subsubsection{Cross-matching known pulsars}
Out of the 21 pulsars detected in the test MWA image (Observation A), 19 of them were detected in a Stokes I RACS image. Of the 2 pulsars not present, one is associated with a supernova remnant, SNR G21.5$-$0.9. A possible reason for the non-detection of the other pulsar, PSR J1822$-$1400, could be very steep spectral index of $-2.25$ \citep[ATNF catalogue]{ref:ATNF}.

\subsubsection{Cross-matching all sources}
\begin{figure}[hbt!]
    \centering
    \includegraphics[width=0.8\textwidth]{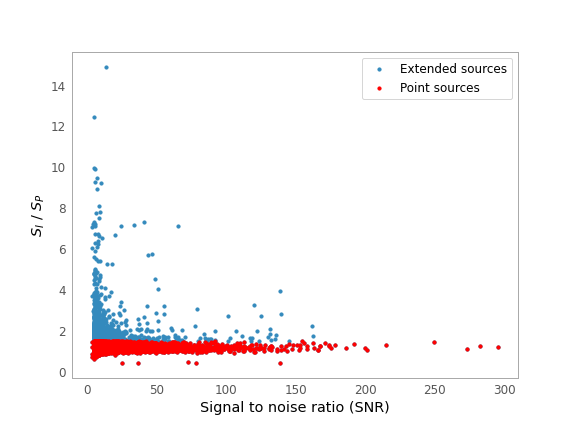}
    \caption{Ratio of integrated flux to peak flux ($\rm S_{\rm I}$ / $\rm S_{\rm P}$) as a function of SNR of the sources. The criterion for the source to be a point source was chosen to be the ratio $\rm S_{\rm I}$ / $\rm S_{\rm P} < 1.5$. Based on this criterion all the point sources are shown in red ($\sim$ 1600) and the extended sources are in blue ($\sim$ 6700).}
    \label{fig:fluxsnr}
\end{figure}

On cross-matching our catalogue with the RACS all-sky catalogue, we find $\sim$ 8000 sources that are present in both the catalogues and 342 sources that are present only in MWA catalogue at 5$\sigma$ confidence level. These 342 sources are of interest to us as potentially intriguing sources for follow up. We further reduce this number by excluding extended sources as pulsars are generally point sources. The ratio of the integrated flux over source in the image ($ S_{ I}$) to peak flux of the source in the image ($ S_{P}$), both calculated by AEGEAN, is used to select the point sources such that sources where $S_{ I} /  S_{ P} > 1.5$ are considered to be extended. This compactness criterion is based on the similar criterion in the GaLactic and Extragalactic All-sky Murchison Widefield Array (GLEAM) catalogue \citep{ref:GLEAM} and is applied to the sources present in both catalogues as well as the sources present in MWA catalogue only. The ratio as a function of the signal to noise ratio (SNR) of the sources calculated using the source catalogue is shown in Figure \ref{fig:fluxsnr}. The blue points ($\sim$ 6700) are the excluded extended sources and the red points ($\sim$ 1600) are the compact sources taken into consideration for further analysis on application of the compactness criterion. This subset of sources is used for further analysis as described in Section \ref{subsec:m}.

\subsection{Testing of the methodologies}
\label{subsec:m}

\subsubsection{Spectral Index}
\label{subsubsec:spec}
Observation A (refer to Table \ref{Table:1}) is predominantly used for the examination of the first criterion. On successful verification of the criterion, it is then independently applied on observations C and D. The process begins by first selecting sources with a steep spectrum, by calculating the spectral index using Equation \ref{equn:spec} for the sources present in both MWA and RACS catalogues. Furthermore, for the sources that are present only in the MWA Stokes I images, an upper limit on the spectral index is calculated using the local $\sigma$ from the RACS noise map (5$\sigma$ for RACS). A recent study of spectral indices of 441 radio pulsars suggests the mean of the spectral index distribution is -1.8 for79\% of pulsars following a simple power law spectrum \citep{ref:Jank}. Taking this into account along with error margins, a spectral index cutoff of -1.2 is selected, such that sources steeper than -1.2 are considered as steep spectrum sources and used for further analysis. While ~29\% of the known pulsar population exhibit spectra flatter than -1.2 \citep{ref:spec}, the steeper cutoff is more likely to find those that were missed by previous surveys at higher frequencies while also reducing false positives from other source types e.g. steep spectra AGN. This cut-off may be relaxed in future (allowing shallower power laws) if additional robust filtering criteria are developed in order to reduce the number of false positives. 

\begin{figure}[hbt!]
    \centering
    \includegraphics[width=\textwidth]{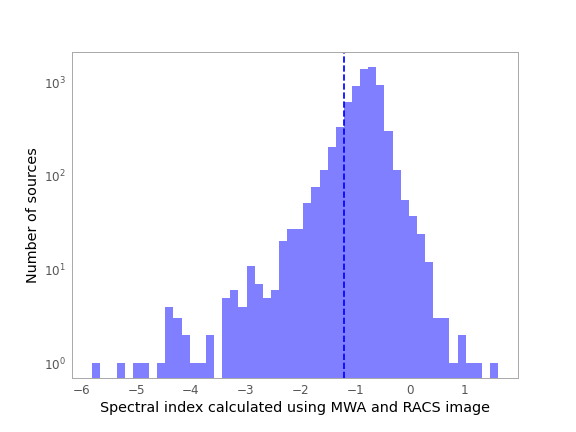}
    \caption{Distribution of the spectral index of the sources detected in the MWA image of observation A. It can be seen that most of the sources have spectral index between -1 and 1. The blue dotted line shows the spectral index cutoff threshold used in this analysis.}
    \label{fig:spec}
\end{figure}

\begin{figure}[hbt!]
    \centering
    \includegraphics[width=\textwidth]{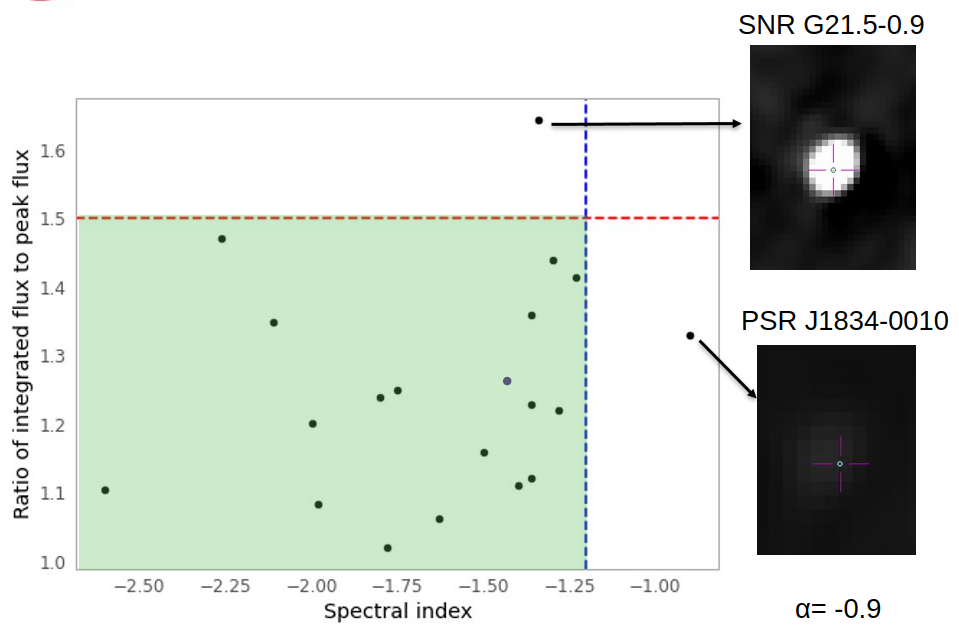}
    \caption{Pulsars that are detected in imaging and the ones that satisfy the criterion of steep spectrum. The blue dashed line shows the spectral index cutoff and the red dashed line signifies the compactness cutoff applied to the sources. The green region shows the parameter space that the combination of the compactness and spectral steepness criteria can probe in imaging domain .}
    \label{fig:pulspec}
\end{figure}

The distribution of the spectral indexes of the sources and the applied cutoff are shown in Figure \ref{fig:spec}. Application of the criterion resulted in a subset of $\sim$ 300 sources that are steep spectrum. Out of these sources, 19 are already known pulsars. Figure \ref{fig:pulspec} shows the pulsars that satisfy the criterion of steep spectrum sources. It also shows the two pulsars that were excluded due to criterion not being satisfied. One is in a supernova remnant and was excluded on application of the compactness criterion. The other pulsar, PSR J1834-0010 has a spectral index of -0.9 which is not steep enough to pass the criterion. The $\sim$ 300 sources that are considered steep according to our criterion were checked for pulsar-like emission by forming a tied-array beam on each of the sources and performing a PRESTO \citep{ref:Ransom} based search up to a DM of 500 pc $\rm cm^{-3}$ with minimum DM steps of 0.02 pc $\rm cm^{-3}$ and maximum of 0.5 pc $\rm cm^{-3}$. No pulsations from these sources were found in the beamformed data. However, as mentioned before the sensitivity of MWA is significantly reduced at DMs above 250 pc $\rm cm^{-3}$. Therefore, deeper searches with a higher DM range with a high-frequency instrument may enable identification of pulses from these sources.

\subsubsection{Circular polarisation}
\label{subsubsec:circpol}

\begin{figure}[hbt!]
    \centering
    \includegraphics[width=\textwidth]{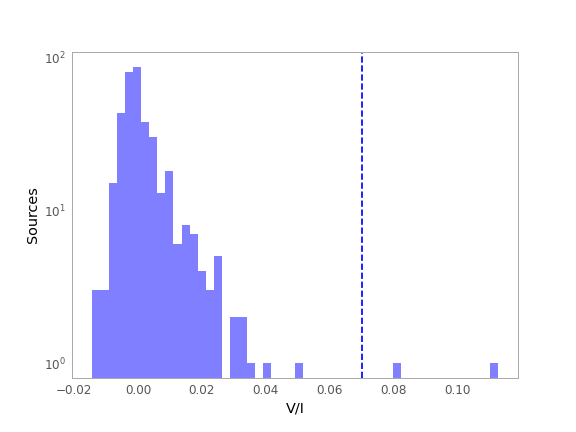}
    \caption{Distribution of fractional polarisation of the sources after the removal of non-physical leakage around the sources. The blue dashed line shows a fractional polarisation threshold of 7$\rm \%$ from the existing literature. The 4 sources with |V/I| > 0.07 are known to be circularly polarised pulsars according to the European Pulsar Network (EPN) database.}
    \label{fig:iv}
\end{figure}

A catalogue of sources from the Stokes V image is generated by running AEGEAN on the Stokes V inverted image and the original Stokes V image. The combination of the catalogues from both the processes is used as the Stokes V source catalogue. The second criterion of circular polarisation uses the sources from Stokes I and V images of observation A to select potential pulsar candidates based on their circular polarisation. Unfortunately, the positive detection of pulsar candidates is hindered by instrumental leakage, even when leveraging the most advanced MWA primary beam model used for this analysis \citep{ref:Soko}. \citet{ref:Lenc2017} showed that such polarisation leakage can be mitigated in drift-scan observations by modelling the leakage pattern across the beam and then subtracting it. A similar approach has been taken in this analysis to deal with the leakage for the observations. For observation A, the Stokes V sources are crossmatched with Stokes I catalogue sources and the median leakage in 5$\degree$ radius around each source is then calculated. This is then subtracted from the measured Stokes V flux density to suppress any residual leakage resulting in a more robust estimate of the source SNR and fractional polarisation. A distribution of fractional polarisation, V/I, of sources in the image for observation A is shown in Figure \ref{fig:iv}. The acceptable threshold for fractional polarisation according to literature \citep{ref:Lenc, ref:Lenc2017} is $\sim$ 7 $\%$. Application of the threshold reveals 4 sources that satisfy this criterion. According to the European Pulsar Network (EPN \href{http://www.epta.eu.org/epndb/}) database, 9 pulsars in this field are circularly polarised. These 4 sources that satisfy our threshold with V/I > 0.07, are a subset of the 9 pulsars that are circularly polarised. Out of the 5 non-detections, one pulsar, PSR J1823-1115, shows a sign reversal in the mean pulse profile of Stokes V polarisation and the other 4 (highest absolute value of 15 mJy/beam) are below our detection threshold in Stokes V image ($\sim$ 35 mJy/beam). No other interesting pulsar candidates were detected in this image using this method. Further investigation in better removing the polarisation leakage is required in order to detect potentially interesting pulsar candidates.

\subsubsection{Variability}
\label{subsubsec:lightcurve}
The previously defined characteristics are not exclusive to pulsars and may be selecting other type of sources for example, flare stars and M-dwarfs \citep{ref:Lynch1, ref:Lynch2}. Even though the luminosities of pulsars are relatively stable when averaged over long time-scales \citep[][]{ref:Taylor,ref:Mat} , the received flux density can be modulated by propagation effects such as refractive and diffractive scintillation, e.g., \citet{ref:Armstrong}. While diffractive scintillation causes variability on time-scales of tens of minutes, refractive scintillation constitutes variations on the time-scales of weeks to months. Depending on the cadence of the observations using the MWA, we explore the variability timescales of 300s, 60s and 30s, probing diffractive interstellar scintillation specifically. This is done by generating light curves of the possible variable sources. These sources are selected based on variability statistics (equations. \ref{equn:3}, \ref{equn:4}, \ref{equn:5}) such as how significant the variability is (significance) and the percentage of variability of (modulation) the pixel (or the source) when compared to its surrounding. The statistical quantities described below are are for a specific image pixel (function of (x,y) position of the image), but this dependence was dropped for brevity.  

For each pixel, the reduced $\chi^{2}$ (significance) statistic is defined as 

\begin{equation}
    \chi^{2} = \sum_{i=1}^{n}\frac{(S_{i} - \bar{S})^{2}}{\sigma_{i}^{2}},
    \label{equn:3}
\end{equation}
where n is the number of images, $S_{i}$ is the flux density of a pixel in image i, $\sigma_{i}^{2}$ is the variance within the images \citep{ref:Bell1}. $\bar{S}$ is the weighted mean flux density defined as

\begin{equation}
    \bar{S} = \frac{\sum_{i=1}^{n}\left ( \frac{S_{i}}{\sigma_{i}^{2}} \right )}{\sum_{i=1}^{n}\left ( \frac{1}{\sigma_{i}^{2}} \right )}
    \label{equn:4}
\end{equation}

\begin{figure}[hbt!]
    \centering
    \includegraphics[width=\textwidth]{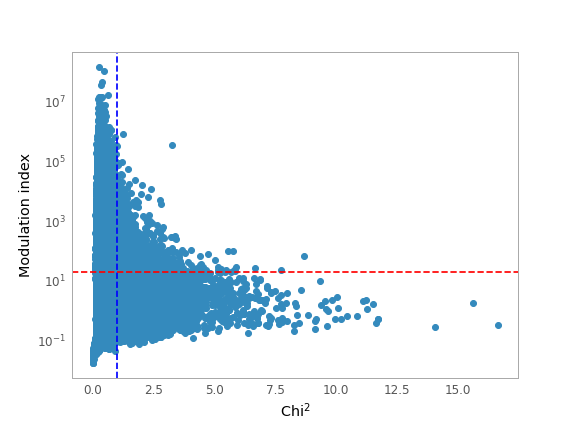}
    \caption{Correlation between $\chi^{2}$ and modulation index for 30sec timescale for observation B. The blue dashed line indicates the $\chi^{2}$ threshold of 1 (for this case). In order to reduce the storage and CPU requirements, these sources are further subjected to a threshold of 20\%, denoted by the red dashed line, for modulation index. The light curves for only sources ($\sim$ 250 sources) satisfying both threshold are saved for further assessment. These light curves are then visually investigated to determine its ranking in terms of variability and searched for pulses.}
    \label{fig:corr}
\end{figure}

In combination with the $\chi^{2}$ test, we also calculate the modulation index of the source/pixel that is used to quantify the degree of variability of the source when compared to its surrounding sources. The modulation index ($m$) is defined as 

\begin{equation}
    m ~ (in ~ \%)= \frac{\sigma_{S}}{\bar{S}}
    \label{equn:5}
\end{equation}
where $\sigma_{S}$ is the standard deviation of the flux density of the pixel and $\bar{S}$ is the mean flux density of the pixel \citep{ref:Bell}. The correlation between $\chi^{2}$ and modulation index as shown in Figure \ref{fig:corr} is used to determine specific threshold for the different timescales for every observation. A threshold is chosen such that the number of candidates to be checked via beamforming does not exceed 1000. For the observations analysed as part of this paper, the thresholds are similar given that it is from the same instrument with similar noise characteristic of the image. For observations which are significantly different, this threshold has to be determined such that the candidate sources to beamform on is not extremely high.

\begin{figure}[hbt!]
    \centering
    \includegraphics[width=\textwidth]{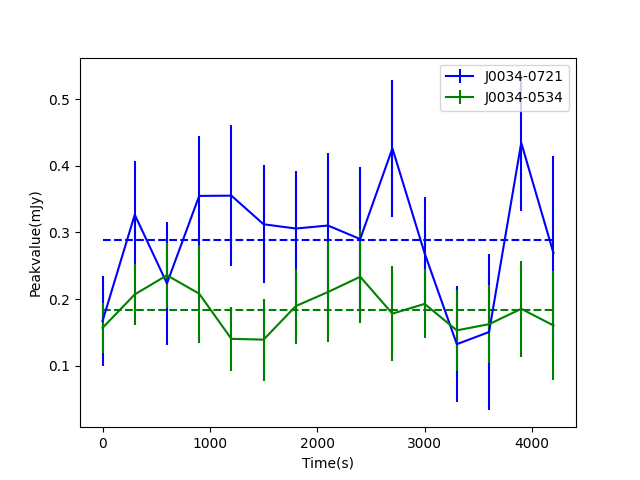}
    \caption{Light curves for PSR J0034-0721 (blue) and PSR J0034-0534 (green) for 300s timescale averaged images for the whole observation. For this test, our $\chi^{2}$ threshold is 1 and the threshold on modulation index (m) is 20\%. The blue light curve has a $\chi^{2}$ of 1.6 and has a modulation index of 30\%, whereas the green curve is approximately constant and has a $\chi^{2}$ of 0.8 and modulation index of 10\%. While the pixel corresponding to the blue curve will be easily selected as a candidate on application of our criterion, the pixel for the green curve will not satisfy our threshold. This supports the variability of J0034-0721 as stated in the literature and indicates that our criterion is indeed useful for detection of true pulsar candidates.}
    \label{fig:0034}
\end{figure}

A test of the above described formalism was done on observation B (see Table \ref{Table:1}) consisting of two known pulsars, PSR J0034-0521 and PSR J0034-0721, the latter of which is variable on short timescales \citep{ref:Mcsweeney}. This test was a sanity check, performed to determine if the variable pulsar would satisfy our criterion without any prior information as one of the variable candidates with high significance. As an initial test case, a $\chi^{2}$ threshold of 1, was applied to the observation, resulting in a list of $\sim$ 5 candidate pixels in a small window around the test pulsar, PSR J0034-0721. Three out of the 5 candidate pixels corresponded to that of the pulsar. The other 2 candidates, when searched using PRESTO did not yield any pulsar like signals. The same analysis was then extended to the whole image, using the same threshold, which resulted in $\sim$ 250 candidates. 
A careful inspection of the light curves reveal one variable source whose pixel position is coincident with that of the variable pulsar PSR J0034-0721. Furthermore, a distinct variability on timescale of 300s is observed for J0034-0721 when compared to the light curve of the pixel position corresponding to that of J0034-0534 as shown in Figure \ref{fig:0034}. The $\chi^{2}$ for this pixel is 1.6 with a modulation index of 30\%, satisfying our threshold of $\chi^{2}$. This demonstrates that any source which is at least as variable as J0034-0721 should be easily detected using the criterion of light curve extraction. The variability procedure is then applied to observations C and D as described in Section \ref{subsec:indi}.

\subsection{Application of the methodologies to other observations}
\label{subsec:indi}
Two observations are used to independently test the formulated methodologies and corresponding criteria to determine the success rate of detecting pulsar candidates. In this section, we describe the results obtained when the test is performed on the observations C and D.

\subsubsection{Observation C}

Processing of this observation was performed as described in Section \ref{subsec:processing} to produce Stokes I and V images. The images produced for this observation were 4096 $\times$ 4096 pixel with 0.5 arcmin/pixel resolution, resulting in ~35\degree $\rm x$ 35\degree images. The $\sigma$ for this observation was 30 mJy/beam (Stokes I) and 10 mJy/beam (Stokes V) and increases almost by a factor of 3 at the edge of the image. AEGEAN \citep[]{ref:Aegean1,ref:Aegean2} was used with a threshold of 5$\sigma$ to extract sources from mean image resulting in $\sim$ 1100 sources in Stokes I image. On the application of the compactness criterion, the extended sources are discarded from further analysis, reducing the number of sources to $\sim$ 800. On performing a  similar comparison to the ATNF catalogue as described in Section \ref{susbec:atnf}, a total of 335 pulsars were present in the 3dB (half power point) beam. We expect to detect 52 pulsars at 5-$\sigma$ level, after inspecting the local standard deviation of the noise near the pulsars. 28 out of the 52 pulsars were detected in imaging. Four of the non detected pulsars were associated with supernova remnants and hence were not in the analysis when the compactness criterion is applied. One possible reason for the remaining non detections was the pulsar flux density being lower than the mean flux density threshold and hence not bright enough for it to be detected in imaging. Another possible reason for the non detection may be the turnover of pulsars, which is more dominant at lower frequencies. On application of the pulsar searching and folding algorithm in PRESTO on the 28 imaging detections, we were able to successfully detect 23 of them. The 5 non-detections can be attributed to the high DM of the pulsars, which significantly reduces the detection sensitivity of periodic searches at MWA frequencies. 

Out of the 28 pulsars detected in imaging, we are able to detect 27 in RACS Stokes I image. The possible reason for the non detection of PSR J1935+2154 could be its association with a supernova remnant, G57.2+0 \citep{ref:snr}.

Out of the compact sources present in our Stokes I catalogue, $\sim$ 500 are present in RACS all-sky catalogue. Around $\sim$ 350 sources are present only in MWA catalogue and are of high significance to us. The spectral index of the sources present in both catalogues is calculated using equation \ref{equn:spec}. The upper limit of the spectral index is calculated for the sources present in only MWA catalogue. The spectral index threshold of -1.2 as established in Section \ref{subsec:m} is applied to both the subsets of sources.  A total of $\sim$ 100 sources have a steep spectrum. 27 of the 100 sources are already known pulsars. The one excluded pulsar has a spectral index of -1.0 and hence does not satisfy our criterion. The remaining steep spectrum sources were beamformed and searched for pulsar-like signal. None were confirmed as pulsar candidates. 

The second criterion was applied to the Stokes I and V catalogues sources in the similar way as described in Section \ref{subsubsec:circpol} after the subtraction of instrumental leakage. When this criterion for selecting polarised sources as defined in Section \ref{subsec:m} was applied, 2 sources are independently detected above the threshold, which can be associated with 2 pulsars. However, the EPN database shows that 4 pulsars should be detected in the Stokes V image. The non-detection of the other 2 pulsars can be attributed to the flux density of the pulsars (highest absolute value is 25 mJy/beam) being below the noise in the Stokes V image ($\sim$40 mJy/beam). No other polarised sources above the threshold were detected in this observation. A reduction of Stokes V leakage will improve the sensitivity of this criterion.  

In order to select sources which are variable in nature and could possibly be interesting pulsar candidates, we generated mean images for timescales of 300s, 60s and 30s. The $\chi^{2}$ vs modulation index maps for three above mentioned timescales is used to set a threshold for $\chi^{2}$ such that we produced light curves for only candidates that are most probable to be pulsar candidates. For this observation, a $\chi^{2}$ threshold of 1, and modulation index threshold of 20\% is applied, resulting in $\sim$ 550 candidates. A careful scrutiny of the resulting light curves was performed and a list of sources and their corresponding positions was created, however, no interesting targets were confirmed as pulsar candidates.   

\subsubsection{Observation D}

Stokes I and V images were again produced following the procedure laid out in Section \ref{subsec:processing}. The images are 8192 $\times$ 8192 pixels with 0.3 arcmin/pixel resolution, creating images that are $\sim$ ~40\degree $\rm x$ 40\degree. The noise in this image varies from 15 mJy/beam at the centre of the images to 30 mJy/beam at the edge of the image in Stokes I. For mean Stokes V image, $\sim$ 2 mJy/beam at the centre of the image and increases by almost a factor of 5 at the edges of the image. The source catalogue for the mean images was produced using AEGEAN and consists of $\sim$ 5000 sources. The removal of the extended sources from the catalogue resulted in $\sim$ 4000 sources. According to the ATNF pulsar catalogue there are 12 pulsars in the field. Ten out of 12 are detected in imaging. The 2 pulsars are not detected due to the local noise of the pulsars being high as both the pulsars are at the edge of the image. 

Around $\sim$ 3800 sources are present in both RACS and MWA catalogue. The spectral index for the sources present in both the catalogues and the upper limit of the spectral index for sources present in only MWA catalogue is calculated using equation \ref{equn:spec}. Application of the spectral index threshold mentioned in Section \ref{subsec:m} results in a total of $\sim$ 250 sources that have steep spectral index. Out of 250 sources, 5 are known pulsars. The rest of the sources are beamformed and searched using the MWA pulsar search pipeline to detect any signal. We were not able to confirm any candidates as pulsars. 

The sources from the Stokes I and V catalogues were then crossmatched and the fractional polarisation is calculated for the sources. After the removal of artifacts and leakage, we apply the fractional polarisation criterion and detect 5 real circularly polarised sources. Out of the 5 circularly polarised sources, Three of them are already known pulsars. The two other sources have been searched via traditional searches but no pulsations have been found. Five pulsars in this observation are expected to be circularly polarised based on EPN database. The 2 non-detected pulsars (highest absolute value is 10 mJy/beam) are below the detection threshold for the Stokes V image for this observation ($\sim$ 32 mJy/beam). 

The variability criterion was then applied at the timescale of 300s, 60s and 30s. A $\chi^{2}$ of 1 and modulation index threshold of 20\% was enforced, resulting in $\sim$ 1000 candidates. All candidates were searched using standard pulsar techniques, but none of the candidates displayed pulsar like emission. 

The independent application of the criterion to the two observations helped in understanding the efficiency of the methodologies in selecting pulsar candidates. It also provided us with a context of the number of sources that one may expect from the different criterion and that increasing thresholds could be one possible way of reducing the number of candidates. 
Table \ref{Table:2} shows the expected and actual pulsar detections when the methodologies are applied to the four observations. The potential reasons for various non detections of known pulsars are stated in the Sections \ref{subsec:m} and \ref{subsec:indi}. 
Table \ref{Table:3} shows the initial number of candidates for the four methodologies when applied to the four observations. It is clear that the circular polarisation criterion produces significantly fewer candidates, which is possibly driven by our lack of full understanding of the polarisation leakage in the MWA images. More effort in understanding this issue and formulating a more robust threshold is required for greater success of the criterion. On the other hand, variability criterion produces a high number of candidates which may result in considerable amount of false positives. One must be careful when determining the threshold for $\chi^{2}$ and modulation index in order to produce a greater number of true pulsar candidates. Lastly, the spectral index criterion, seems to generate a manageable number of candidates that can be followed up using periodicity or single-pulse searches while being computationally inexpensive. 

   \begin{table*}
      \centering

    $$
        \begin{tabular}{ *{10}{c} }
    \hline
\textbf{Obs Name}    & \multicolumn{2}{c}{\textbf{Pulsars}} 
            & \multicolumn{2}{c}{\textbf{Spectral Index}}
                    & \multicolumn{2}{c}{\textbf{Polarisation}}
                            & \multicolumn{2}{c}{\textbf{Variability}}                \\
    \hline
    &   Expected & Actual & Expected & Actual & Expected & Actual & Expected & Actual \\
    \hline
A   &   42  &   21  &   42  &   18  &  9   &   4  &   N/A  &   N/A  \\
    \hline
B   &   10  &   8  &  10  &   7  &   2  &   1  &   1  &   1  \\
    \hline
C   &   52    &   28    &   52    &   27    &    4   &    2   &    N/A   &   N/A    \\
    \hline
D   &   12    &    10   &    12   &    5   &   5    &    3   &   N/A    &    N/A   \\
    \hline
\end{tabular}
    $$
    \caption[]{Table shows the expected vs actual pulsar detections on application of the methodologies described in the paper. The last columns of observation A, C and D are empty as there are no variable pulsars in the field and hence the criterion of variability is not applicable for these observations.}
    \label{Table:2}
    \end{table*}
%

   \begin{table*}
      \centering

    $$
        \begin{tabular}{ccccccc}
        \hline
\textbf{OBS Name} & \textbf{Observation ID} & \textbf{Spectral Index} & \textbf{Circular polarisation} & \textbf{Variability}   & \textbf{Total number} & \textbf{Candidates after}                   \\
        
&&&&&\textbf{of candidates}& \textbf{combining criteria}\\
    \hline
A & 1276619416 & 300  & 1  &  150 & 451  &7\\
B & 1150234552 & 150  &  1 &  250 & 401  &4\\
C & 1148063920 & 100  &  2 &  550 & 652  &5\\
D & 1274143152 & 250  &  5 &  1000& 1255  &4\\
        \hline
    \end{tabular}
    $$
    \caption[]{Table shows the total number of candidates when the four methodologies have been applied to the 4 observations. The last column shows the final candidates after combining the criteria like spectral index and circular polarisation and spectral index and variability.}
    \label{Table:3}
    \end{table*}

\section{Discussion}

Even though there are lots of pulsars in the field of view of the observations, the number of pulsar detection in imaging is affected by the sensitivity of the images, reducing the number of pulsar detections in the imaging space. We also performed a comparison of the number of pulsars that were detected in imaging space and the detections of the same pulsars when searched for periodic signals using PRESTO. We can clearly deduce that the high DM pulsars are not detected when we perform a periodic search. For example, in observation A, PSR J1835-0643, was detected at $\sim$ 5$\sigma$ in Stokes I image, but was not detected when a periodic search was performed due to its high DM of $\sim$ 473 pc $\rm cm^{-3}$. Another reason for the non detection of the pulsars in periodic searches is the effect of scintillation and scattering. For example, PSR J1833-0338, in observation A Stokes I image was detected at $\sim$10$\sigma$ via imaging as well as periodic searches. The expected detection significance for the pulsar is $\sim$ 30$\sigma$ for periodic searches based on its flux density. Due to the scattering timescale of the pulsar ($\sim$ 0.7s) being close to its period ($\sim$ 0.6s), it suffers from scattering which degrades the detection sensitivity of the pulsar when searching using PRESTO and was hence was detected at a lower significance in periodic searches. On the other hand, there are pulsars such as PSR J1801-0357 that are detected by periodicity searches and not in image-based searches due to its flux density being lower than the flux density detection threshold for imaging. Overall, image-based searches target pulsar candidates that may have been missed by traditional searches due to scattering or high DM. Any such pulsar candidates need to be followed up with high frequency telescopes such as the Parkes radio telescope (Murriyang) or uGMRT. 

Another advantage of image-based pulsar candidate searches over traditional pulsars searches is the notably reduced time and resources that image based searches take to produce a list of pulsar candidates. If we perform a periodicity search for every pixel in our image, i.e. 10 million pixels, it would take 40 million CPU hours to obtain pulsar candidates. However, on applying the methodologies, we reduce the number of follow-up processing steps to a more tractable $\sim$ 500 per observation.

\subsection{Criterion 1 - steep spectrum}
The first criterion focuses on candidates that have spectral index $< -1.2$. We were able to successfully detect most of the pulsars in the Stokes I images which satisfy our criterion. However, we did not detect any new pulsar candidates via this method. The steep spectrum sources detected in this criterion may be other categories of extragalactic sources, such as extended emissions from radio halos and relics in merging galaxy clusters and do not have a pulsar like signal and hence not detected via PRESTO searches. It may happen that sources that are not detected as pulsars via PRESTO-based searches are indeed pulsars but not detected due to high scattering and hence cannot be confirmed by using only the MWA data. Such candidates needs to be observed with telescopes operating at higher frequencies. Overall, selecting an ideal upper limit of the spectral index that increases the fraction of pulsars detected without any prior bias while minimising the fraction of false detections of background radio sources is essential for success of this criterion. In this paper, we have used a spectral index threshold of -1.2 for selection of sources. By doing so, we were able to detect $\sim$ 90\% of the pulsars detected via imaging in all the observations along with an average of $\sim$ 200 new pulsar candidates that are steep spectrum for each observation. Doing similar analysis with a higher threshold, for example, $\alpha = -3$, would result in significantly less candidates but at the expense of detecting fewer known pulsars. Ideally, a threshold that is able to redetect the maximum number of pulsars as well as generate a manageable number of candidates is preferred, hence, our threshold of  $\alpha < -1.2$, seems appropriate for the future processing of much larger set of MWA observations. 

\subsection{Criterion 2 - circular polarisation}
We also examined the Stokes I and V images to detect candidates that may be circularly polarised. Around 50\% of the circularly polarised pulsars in the observation were detected as candidates without any prior knowledge about their polarisation. However, we were not able to detect any potential pulsar candidate with high significance via this method when the fractional polarisation threshold was applied. Compared to total intensity, the number of candidates resulting from the application of this criterion is lower by almost an order of magnitude. Selection of an appropriate threshold for candidates to be considered circularly polarised is beneficial to detect genuine Stoke V polarised sources and pulsar candidates. Moreover, any candidate that is circularly polarised but not a pulsar is likely to be objects of scientific interest. For dipole based instruments such as the MWA, the instrumental leakage, where emission from Stokes I leak into other Stokes parameters has to be better understood. This leakage significantly affects the detection of circularly polarised pulsars and increases the number of false detections. Even with the improvements in calibration and the model of the primary beam, errors are still be present due to the imperfections of the model and the MWA dipoles itself. In Section \ref{subsec:m} we describe the process of fitting a surface to the fractional polarisation of the sources for the observations, in order to account for the leakage and detect true pulsar candidates. However, further reduction of the leakage will be necessary to reduce the threshold and increase sensitivity of this criterion.

\subsection{Criterion 3 - Variability}
Another aspect of pulsar candidate selection tested is by taking advantage of the variability of the sources in different timescales. This method mainly focuses on detecting pulsar candidates that may be missed by traditional searches due to their nulling behaviour or scintillation. We were able to detect the variable pulsar, PSR $\rm J0034-0721$ in the test observation as a pulsar candidate with high significance. Extracting light curves for every single pixel of the image will result in $\sim$ 1 million candidates, which is a large number of candidates per image to follow up. In order to limit the number of candidates to a few hundreds, we apply a threshold of $\chi^{2}$ and modulation index as described in Section \ref{subsec:m} by analysing the correlation between $\chi^{2}$ and modulation index for every timescale we process. Even with those reductions in place, it is still not feasible to follow up on all the candidates with other telescopes. The benefit of leveraging the MWA VCS data means that all of them can be searched using products from the same telescope and observation. The other alternative is to check if these candidates satisfy the other two criteria, and ranking the candidates according to the number of criteria satisfied. This set of candidates ( $\sim$ 10) can then be followed up with other telescopes.  
Unfortunately, we were not able to detect any pulsar-like candidates. One of the reasons for the non-detections could be that the sources are variable on longer timescales than the ones that are chosen for the observations. It may also be that the variable sources are not pulsars but part of other variable populations such as flare stars. The lack of detection could also be related to the constraints on our periodicity search using PRESTO at MWA frequencies. It should be noted an improved understanding of imaging artifacts produced while imaging with the MWA is required in order to distinguish between true variable sources and sources that appear to be variable but are in fact spurious detections.

\section{Conclusions and Future Work}

Overall, the three methods have been successfully curated and tested on the observations given in Table \ref{Table:1}. We have a detection efficiency of 50\% for both method I (steep spectrum) and II (circular polarisation), while method III (variability) was tested on a small number of test sources all of which were successfully retrieved, after the relevant pulsars for each observation were taken into account. A thorough analysis of the candidates resulting from the application of the above methodologies to the observations was performed. All candidates were searched using MWA beamformed data, however we were not able to detect any pulsar signal. The larger number of initial candidates ($\sim$ 100) were reduced by combining the different methodologies to produce a list of 20 candidates, which can then be followed up with higher frequency telescopes. Care must be taken while combining the criteria as we may be missing interesting sources that may not necessarily satisfy more than one criteria. \\
Another way to reduce the number of candidates can be optimisation of the thresholds for all the criteria. This is an ongoing project and will be informed by processing a larger number of MWA VCS observations. Overall, it is demonstrated that we were able to detect a large fraction (> 50 \%) of the pulsars expected to be found as well as a selection of new pulsar candidates for the observations.  It is also important to understand the instrumental beam response of the instrument being used as any errors in the beam model will result in leakage. For example, effects such as "beam squint" affect radio telescopes with off-axis feeds such as the (E)VLA \citep{ref:Uson}, and may also be relevant to the future low-frequency telescopes such as SKA-Low. This is useful in informing pulsar search methodologies with the MWA, and suggests that the SKA-Low will be a powerful telescope to find pulsar candidates in continuum images.

These three image-based detection methods outlined here will be applied to $\sim$ 50 MWA observations in an attempt to make the first image-based detection of pulsar with the MWA. The success of this pilot project demonstrates the potential of using image-based methodologies to detect pulsar candidates at low frequencies and improve the efficiency of future pulsar detections with the MWA as well as the SKA-Low.

\begin{acknowledgement}

This scientific work makes use of the Murchison Radio-astronomy Observatory, operated by CSIRO. We acknowledge the Wajarri Yamatji people as the traditional owners of the Observatory site. Support for the operation of the MWA is provided by the Australian Government (NCRIS), under a contract to Curtin University administered by Astronomy Australia Limited. This work was further supported by resources provided by the Pawsey Supercomputing Centre with funding from the Australian Government and the Government of Western Australia. Part of this research has made use of the EPN Database of Pulsar Profiles maintained by the University of Manchester. S.S. acknowledge the support from
Australian Government Research Training Program Scholarship.
\end{acknowledgement}

\bibliography{main}

\end{document}